\documentclass[%
 aip,
 jcp,
 amsmath,amssymb,
 reprint,%
 floatfix,
]{revtex4-2}
\usepackage[utf8]{inputenc}
\usepackage{amsmath}
\usepackage{mathtools}
\usepackage{graphicx}   
\usepackage{bm}
\usepackage{hyperref}
\usepackage{color}
\usepackage{array}
\usepackage{siunitx}
\usepackage[version=4]{mhchem}
\usepackage[normalem]{ulem}
\usepackage{lipsum}
\usepackage{comment}
\graphicspath{{./}{../graphics/}{../graphics/plots/}}

\newcommand{\no}{\nonumber}
\newcommand{\eqn}[1]{Eq.~\eqref{#1}}

\newcommand{\fig}[1]{Fig.~\ref{#1}}
\newcommand{\Fig}[1]{Figure~\ref{#1}}
\newcommand{\Sec}[1]{Section~\ref{#1}}
\renewcommand{\sec}[1]{Sec.~\ref{#1}}

\newcommand{\Refx}[1]{Ref.~\onlinecite{#1}}

\newcommand{\kB}{k_\mathrm{B}}

\newcommand{\pder}[3][]{\frac{\partial^{#1}{#2}}{\partial{#3}^{#1}}}

\newcommand{\wns}{\ensuremath{\mathrm{cm}^{-1}}}

\newcommand{\cvec}[1]{\ensuremath{\mathbf{#1}}}
\renewcommand{\vec}[1]{\ensuremath{\mathbf{#1}}}

\newcommand{\com}{\ensuremath{\mathrm{com}}}
\newcommand{\qmf}{\ensuremath{\overline{F}}}
\newcommand{\qcq}{\ensuremath{\overline{Q}}}
\newcommand{\qcqvec}{\ensuremath{\overline{\vec{Q}}}}
\newcommand{\qcpvec}{\ensuremath{\overline{\vec{P}}}}
\newcommand{\qfint}{\ensuremath{\overline{\vec{f}}_{\mathrm{int}}}}
\newcommand{\qfrot}{\ensuremath{\overline{\vec{f}}_{\mathrm{rot}}}}
\newcommand{\qctau}{\ensuremath{\overline{\bm{\tau}}}}
\newcommand{\qftr}{\ensuremath{\overline{\vec{f}}_{\mathrm{trans}}}}

\begin{document}

\title{Improved torque estimator for condensed-phase quasicentroid molecular dynamics}

\author{George Trenins}%
\email{georgijs.trenins@phys.chem.ethz.ch}
\affiliation{Laboratory of Physical Chemistry, ETH Z\"{u}rich, 8093 Z\"{u}rich, Switzerland.}

\author{Christopher Haggard}%
\affiliation{ 
Yusuf Hamied Department of Chemistry, University of Cambridge, Lensfield Road,\\ Cambridge,
CB2 1EW, UK.%
}%

\author{Stuart C.\ Althorpe}%
\email{sca10@cam.ac.uk}
\affiliation{ 
Yusuf Hamied Department of Chemistry, University of Cambridge, Lensfield Road,\\ Cambridge,
CB2 1EW, UK.%
}%

\date{\today}

\begin{abstract}
We describe improvements to the quasicentroid molecular dynamics (QCMD) path-integral method, which was developed recently for computing the infrared spectra of condensed-phase systems. The main development is an improved estimator for the intermolecular torque on the quasicentroid. When applied to \mbox{qTIP4P/F} liquid water and ice, the new estimator is found to remove an
artificial $25$ cm$^{-1}$ red shift from the libration bands, to increase slightly the intensity of the OH stretch band in the liquid, and to reduce small errors noted previously in the QCMD radial distribution functions. We also modify the mass-scaling used in the adiabatic QCMD algorithm, which allows the molecular dynamics timestep to be quadrupled, thus reducing the expense of a QCMD calculation to twice that of Cartesian centroid molecular dynamics for  \mbox{qTIP4P/F} liquid water at 300 K, and eight times for ice at 150 K.
\end{abstract}

\maketitle

\section{\label{sec:intro}%
Introduction}

Quasicentroid molecular dynamics (QCMD) is a recently developed path-integral dynamics method\cite{Trenins2019} which has given  promising results for simulations of infrared spectra in gas and condensed phases.\cite{Trenins2019,Benson2019,Haggard2021,Fletcher2021} It avoids most of the 
well-characterised drawbacks\cite{Habershon2008,Witt2009,Ivanov2010,Rossi2014,Trenins2018,
Rossi2018}
of the related centroid molecular dynamics (CMD)\cite{Cao1994b,Hone2004,Hone2006} and (thermostatted) ring-polymer molecular dynamics \mbox{[(T)RPMD]}\cite{Rossi2014,Rossi2018,Craig2004,Habershon2012,MillerThomasF2005b} methods,
by propagating classical trajectories on the potential of mean force (PMF) obtained
by constraining a set of curvilinear centroids. These coordinates are
system-dependent and chosen to give compact ring-polymer distributions
in which the curvilinear centroid lies close to the Cartesian centroid
and is hence referred to as the `quasicentroid'.  
As originally developed, QCMD uses a modification\cite{Trenins2019}
of the adiabatic CMD propagator\cite{Hone2004,Hone2006}
which makes QCMD more expensive than CMD (e.g., by a factor of 32 for 
\mbox{qTIP4P/F}\cite{Habershon2009} ice at 150~K), and thus much more expensive than \mbox{(T)RPMD}. However, exciting recent progress has been made in developing a fast QCMD (f-QCMD) 
algorithm,\cite{Fletcher2021} which extracts a highly accurate approximation to the
quasicentroid PMF from a static path-integral calculation.\footnote{To date, f-QCMD
has been applied only in the gas-phase, but condensed-phase applications are likely
soon.}

\footnotetext[2]{For gas-phase water and ammonia the intensities of the overtone and combination bands can be corrected using harmonic perturbation theory;\cite{Ple2021,Benson2021} for methane this approach has proved less successful.\cite{Fletcher2021}}
\setcounter{footnote}{2}
  
In the gas-phase, QCMD has now been applied to water, ammonia and methane (the latter using f-QCMD).\cite{Trenins2019,Benson2019,Haggard2021,Fletcher2021}
The positions and intensities of the  fundamental bands are in excellent agreement
with the exact quantum results, and the positions of the overtone and combination
bands are in reasonable agreement. However, the intensities of the latter are an
order of magnitude too small, which was shown
recently\cite{Ple2021,Benson2021} to be because QCMD neglects coupling between 
the Matsubara dynamics\cite{Hele2015,Hele2015a,Trenins2018,Jung2019, Althorpe2021}
of the centroid and the fluctuation modes [as also do CMD, RPMD and classical molecular dynamics (MD)].\cite{Note2}

In the condensed phase, QCMD has been tested on the \mbox{qTIP4P/F} water model, for the liquid at 300 K and for ice I$_\text{h}$ at 150 K.\cite{Trenins2019,Benson2019}
Detailed comparisons with experiment are not possible for such a simple potential,\footnote{QCMD calculations using more realistic water potentials such as the MB-pol surface of \Refx{Babin2014} or the DFT scheme of \Refx{Marsalek2017} have not yet been reported.} but the QCMD results are promising: the spectra appear to inherit the advantages of the gas-phase QCMD calculations, with the stretch bands showing none of the artefacts that appear in the corresponding CMD and (T)RPMD spectra at low temperatures\cite{Rossi2014a} and lining up well with the results of other simulation methods.\cite{Liu2015,Liu2011a} 
 
However, QCMD does appear to have introduced a minor artefact of its own into the \mbox{qTIP4P/F} spectra, in the form of a 25 cm$^{-1}$ red shift in the libration bands of both the liquid and ice.\cite{Trenins2019,Benson2019} 
This is a numerically small error, but it might become larger in other condensed-phase calculations,  because it is
most likely caused by errors in the estimator for the intermolecular torque $\qctau$ (between the quasicentroids of pairs of monomers). Unlike the other components of the PMF, $\qctau$ needed to be approximated in order to make the simulations practical, and 
the ad hoc assumption was made\cite{Trenins2019} that $\qctau$ could be approximated as the average of the torques on the individual ring-polymer beads. 

Here, we introduce a new estimator for $\qctau$ and report numerical tests which show that it removes the 25 cm$^{-1}$ red-shift from the \mbox{qTIP4P/F} libration bands, increases slightly the intensity of the stretch band in the liquid, and reduces small errors in the radial distribution functions (noted in previous QCMD calculations at 300~K\cite{Trenins2019}). After summarising the theory of QCMD in \sec{sec:backgr}, we introduce the new $\qctau$ estimator in \sec{sec:torque}, and report the tests on \mbox{qTIP4P/F} water in \sec{sec:simulations}. We also (\sec{ssec:m-weighting}) introduce a straightforward modification to the adiabatic QCMD propagator which improves numerical stability and speeds up the calculation by a factor of four. \Sec{sec:conclusions} concludes the article.

\section{\label{sec:backgr}%
Background theory}

In QCMD, the dynamics are approximated by classical trajectories on a quantum potential of mean force,  which is obtained by sampling the quantum Boltzmann distribution as a function of a set of curvilinear centroid coordinates. The usual Cartesian `ring-polymers' \cite{Chandler1981, Parrinello1984} are used to represent the distribution.

 For a gas-phase system, the distribution takes the form ${\cal{N}}^{-1}\exp[-\beta W({\bf q})]$,  in which ${\cal{N}}$ is a normalisation factor,
$\beta = 1/\kB T$, and
\begin{subequations}
\label{allw}
\begin{gather}
\label{eq:pimd-pot}
	W(\cvec{q}) = U(\cvec{q}) + S(\cvec{q}), \\
\label{eq:mats-pot}
	U(\cvec{q}) = \frac{1}{N} \sum_{i=1}^{N} V(\cvec{q}_i), \\
\label{eq:spring-pot}
	S(\cvec{q}) = \frac{N}{2 (\beta \hbar)^2} \sum_{\alpha=1}^{n} 
	\sum_{i=1}^{N} m_{\alpha} \big\lVert \vec{q}^{(\alpha)}_{i+1} - \vec{q}^{(\alpha)}_{i} \big\rVert^2,
\end{gather}
\end{subequations}
where $V$ is the system potential, $n$ is the number of atoms, 
$m_{\alpha}$ is the mass of atom $\alpha$,
$\vec{q}^{(\alpha)}_{\mkern3mu i} \! \equiv \big(  
q^{(\alpha)}_{1,i},\,%
q^{(\alpha)}_{2,i},\,%
q^{(\alpha)}_{3,i}%
\big)$, and $q_{\mathrlap{\mkern2mu\rule{0pt}{1.5ex}\nu}}{}^{(\alpha)}$
 are a set of $N$ replica `beads' of the Cartesian  coordinates  ($\nu$=1,2,3 corresponding to $x,y,z$) of atom $\alpha$.

The curvilinear centroid coordinates are system-dependent and are chosen to make the distribution compact, such that the molecular geometry  specified by the curvilinear centroids, called the `quasicentroid', is close to the geometry specified by the (Cartesian) centroids of the atoms (i.e., the centres of mass of $\vec{q}^{(\alpha)}_{\mkern3mu i}$). For example, a good set of curvilinear centroids for gas-phase
 water are the bond-angle centroids 
\begin{equation}\label{polar}
\begin{aligned}
R_{1,2} & =\frac{1}{N}\sum_i r_{i1,2},\\
\Theta & = \frac{1}{N}\sum_i \theta_i,
\end{aligned}
\end{equation}
where $(r_{i1}, r_{i2}, \theta_i)$ are the bond-angle coordinates of the individual beads.  To generate the dynamics, the quasicentroid coordinates are converted to a set of cartesians,\footnote{This does not give back the Cartesian centroids owing to the non-linearity of the Cartesian to bond-angle coordinate transformation.} denoted ${\qcqvec}^{(\alpha)}$. One then propagates ${\qcqvec}^{(\alpha)}$ and the conjugate momenta ${\qcpvec}^{(\alpha)}$ using standard Cartesian classical dynamics, with the forces generated by the PMF
\begin{equation}
\label{eq:qc-force}
{-\pder{\qmf}{\qcqvec^{(\alpha)}}} \simeq {-\left \langle \pder{U}{\qcqvec^{(\alpha)}} \right\rangle,}
\end{equation}
 where  
$\langle \ldots \rangle$ denotes a quasicentroid-constrained average over the ring-polymer distribution. The trajectories are thermostatted, which ensures that they sample a good approximation to the exact quantum Boltzmann distribution, provided the quasicentroid-constrained distributions are sufficiently compact (which should always be checked numerically by comparing static thermal averages computed using QCMD and standard path-integral methods).

For a condensed-phase system, the ring-polymer distribution of \eqn{allw} generalises in the usual way, to include a sum over all the molecules in the simulation cell, together with intermolecular components of the physical potential
$ V$ (including the usual Ewald-sum terms\cite{AllenBook} to handle the periodic boundary conditions). In addition to
{\em internal} components of the quasicentroid (e.g., \eqn{polar} in the case of water),
one must also  define {\em external} components, to specify the centres of mass and orientations of the molecules. 
This is done by applying to the quasicentroids of each molecule the `Eckart-like' conditions,\cite{Eckart1935,WilsonBook}
\begin{subequations}
\label{eq:eck}
\begin{align}
\label{eq:eck-trans}
\sum_{\alpha} m_{\alpha} (\vec{Q}^{(\alpha)} - \qcqvec^{(\alpha)}) & = \vec{0}, \\
\label{eq:eck-rot}
\sum_{\alpha} m_{\alpha} \qcqvec^{(\alpha)} \! \times (\vec{Q}^{(\alpha)} - \qcqvec^{(\alpha)}) & = \vec{0},
\end{align}
\end{subequations}
where the sum is over the atoms in the molecule, and
\begin{equation}
\vec{Q}^{(\alpha)} = \frac{1}{N} \sum_{i=1}^{N} \vec{q}^{(\alpha)}_i
\end{equation}
are the Cartesian centroids of atom $\alpha$. Equation (\ref{eq:eck-trans}) constrains the quasicentroid centre of mass 
\begin{equation}
\qcqvec_{\com} \equiv 
\frac{ \sum_{\alpha} m_{\alpha} \qcqvec^{(\alpha)} }{%
\sum_{\alpha} m_{\alpha}} 
\end{equation}
to lie at the centroid centre of mass (i.e., the overall centre of mass of all the atomic bead coordinates in the molecule).
Equation (\ref{eq:eck-rot}) orients the atomic quasicentroids so as to minimise the mass-weighted sum of their square distances from the atomic quasicentroids ${\qcqvec}^{(\alpha)}$ (see Fig.~6 of \Refx{Trenins2019}).
The potential gradient in the PMF then takes the form
\begin{equation}\label{boris}
{-\pder{U}{\qcqvec{}^{\mathrlap{(\alpha)}\ }}\mkern8mu } = \qfint + \qftr + \qfrot,
\end{equation}
where $\qfint$ denotes the forces acting on the internal components of the centroid (e.g., $R_{1,2},\Theta$ in the case of water),  $\qftr$ denotes the forces between the monomer centres of mass, and  $\qfrot$  the torques between pairs of monomers.

 In \Refx{Trenins2019}, it is shown that $\qfint$ and $\qftr$ can be evaluated directly, but that $\qfrot$ must be approximated. This is because 
 $\qfrot$ is given by
\begin{equation}
\label{eq:qfrot}
\qfrot^{(\alpha)} \equiv m_{\alpha} (\vec{I}^{-1} \qctau) \times \qcqvec^{(\alpha)},
\end{equation}
in which $\vec{I}$,  the inertia tensor with elements
\begin{equation}
I_{\mu \nu} = \sum_{\alpha} m_{\alpha} \!\left[ \Big \lVert
\qcqvec{}^{(\alpha)}
\Big \rVert^{\!2} \delta_{\mu \nu} - 
\qcq^{(\alpha)}_{\mu} \qcq^{(\alpha)}_{\nu}
\right]\!,
\end{equation}
can be be evaluated directly, but the  quasicentroid torque,
\begin{equation}\label{bug}
\qctau=-\sum_\alpha\left(\qcqvec{}^{(\alpha)} -\qcqvec_{\com}\right)\times \pder{U{\phantom{{}^{(\alpha)}}}}{\qcqvec{}^{(\alpha)}},
\end{equation}
cannot be, because it depends on all the components of ${\partial U}/\partial{\qcqvec{}^{(\alpha)}}$. \footnote{Derivatives with respect to the internal quasicentroid  coordinates (e.g., $R_{1,2}$ and $\Theta$) are easily expressed in terms of the Cartesian bead coordinates, as are the derivatives with respect to the molecular centres of mass. However application of the chain rule to the remaining external quasicentroid coordinates would require one to derive a $3 n (N-1)$-dimensional set of curvilinear coordinates orthogonal to $\qcqvec{}^{(\alpha)}$.} Reference~\onlinecite{Trenins2019} therefore introduced the ad hoc approximation
\begin{equation}
\label{eqn:old_estimator}
\qctau \simeq \frac{1}{N}\sum_i {\bm \tau}_i
\end{equation}
where  ${\bm \tau}_i$ are the individual torques on the polymer beads. As mentioned in the Introduction, this approximation works well on the whole but is thought to be responsible for the 25 cm$^{-1}$ red shifts in the libration bands of the spectra of 
\mbox{qTIP4P/F} water and ice, as well as for small but noticeable discrepancies between the QCMD and PIMD radial distribution functions (RDFs).

\section{\label{sec:torque} Improved torque estimator}

We now propose a new estimator for $\qctau$ which is based on a heuristic approximation to the following 
(exact) linear equations for $\vec{I}^{-1} \qctau$, 
  \begin{align}
\label{eq:torque}
&\sum_{\alpha, \nu} m_{\alpha} \Big\{
\qcq{}_{\mu}^{(\alpha)}  Q{}_{\nu}^{(\alpha)}
- \big( \qcqvec{}^{(\alpha)} \! \cdot \vec{Q}{}^{(\alpha)}
\big)
\delta_{\mu \nu}
\Big\} \Big\{
\vec{I}^{-1} \qctau
\Big\}_{\nu} \no \\
& \qquad {} = \sum_{\alpha} \left\{
{-\pder{U}{\qcqvec^{\mathrlap{(\alpha)}\ }}\mkern8mu }
\times \vec{Q}{}^{(\alpha)} - \qfint \times  \vec{Q}{}^{(\alpha)}
\right\}_{\!\mu}
\end{align}
obtained by applying $\times \vec{Q}{}^{(\alpha)}$ to both sides of \eqn{boris}. 
These equations contain the difficult-to-evaluate derivatives ${\partial U}/\partial{\qcqvec{}^{(\alpha)}}$, but now crossed with the Cartesian centroid. In what follows, we give a heuristic justification for replacing this term by 
its easy-to-evaluate\footnote{Derivatives with respect to the Cartesian centroids are easily expressed in terms of bead coordinates.} `complement' ${\partial U}/\partial{\vec{Q}^{(\alpha)}}\times\qcqvec{}^{(\alpha)}$, giving
  \begin{align}
\label{tarquin}
&\sum_{\alpha, \nu} m_{\alpha} \Big\{
\qcq{}_{\mu}^{(\alpha)}  Q{}_{\nu}^{(\alpha)}
- \big( \qcqvec{}^{(\alpha)} \! \cdot \vec{Q}{}^{(\alpha)}
\big)
\delta_{\mu \nu}
\Big\} \Big\{
\vec{I}^{-1} \qctau
\Big\}_{\nu} \no \\
& \qquad {} \simeq \sum_{\alpha} \left\{
{-\pder{U}{\vec{Q}^{\mathrlap{(\alpha)}\ }}\mkern8mu }
\times \qcqvec{}^{(\alpha)} - \qfint \times  \vec{Q}{}^{(\alpha)}
\right\}_{\!\mu}.
\end{align}
 These equations are cheap numerically to set up and solve at every timestep of the propagation, yielding $\vec{I}^{-1} \qctau$ and thus
 $\qfrot$.

We justify this approximation by noting  that the quasicentroid-constrained
 distributions are expected to be compact, such that $\qcqvec{}^{(\alpha)}$ is close to $\vec{Q}^{(\alpha)}$, and that
  the QCMD dynamics is expected to give a good approximation to the Matsubara dynamics of the centroids.\footnote{
 The ensembles of Matsubara trajectories that survive the quasicentroid-constrained Boltzmann averaging
   are thus expected to be as compact as the quasicentroid ring-polymer distributions.} All the normal modes of the ring-polymers evolve explicitly in time in Matsubara dynamics, which allows us to write out the second time-derivative of \eqn{eq:eck-rot} in the form
\begin{align}\label{putin}
\sum_{\alpha}  m_{\alpha} \ddot{\qcqvec}{}^{(\alpha)} \times \vec{Q}^{(\alpha)} & = {} \\
 \sum_{\alpha}  m_{\alpha} \ddot{\vec{Q}}^{(\alpha)} \times \qcqvec{}^{(\alpha)}
& + 2 \sum_{\alpha}m_{\alpha}\dot {\vec{Q}}^{(\alpha)}\times\dot{\qcqvec}{}^{(\alpha)}, \no
\end{align}
 then to average over the quasicentroid-constrained quantum Boltzmann distribution, to obtain
\begin{align}
\sum_{\alpha}  \left \langle m_{\alpha} \ddot{\qcqvec}{}^{(\alpha)} \times  \vec{Q}^{(\alpha)}\right \rangle =& 
\sum_{\alpha} \left \langle m_{\alpha} \ddot{\vec{Q}}^{(\alpha)} \times \qcqvec{}^{(\alpha)} \right \rangle.
\end{align}
The final term in \eqn{putin} has vanished under thermal averaging, since the component of the Cartesian centroid momentum $m_{\alpha}\dot {\vec{Q}}^{(\alpha)}$ that is orthogonal to $\dot{\qcqvec}{}^{(\alpha)}$ is isotropically distributed. 
Noting that forces in Matsubara dynamics are simply the negative derivatives of $U$ (and assuming that the quasicentroid curvature is sufficiently small that $m_{\alpha} \ddot{\qcqvec}{}^{(\alpha)} \simeq \dot {\qcpvec}{}^{(\alpha)} $), we can then write
\begin{align}\label{keir}
\sum_{\alpha}  \left \langle {-\pder{U}{\qcqvec{}^{\mathrlap{(\alpha)}\ }}\mkern8mu } \times  \vec{Q}^{(\alpha)}\right \rangle \simeq& 
\sum_{\alpha} \left \langle {-\pder{U}{\vec{Q}^{\mathrlap{(\alpha)}\ }}\mkern8mu }\times \qcqvec{}^{(\alpha)}\right \rangle .
\end{align}
Finally, we note that the PMF requires only the quasi-centroid-constrained average $\left\langle \vec{I}^{-1} \qctau\right\rangle$, 
and that \eqn{eq:torque} implies that
 \begin{align}
\label{resign}
&\sum_{\alpha, \nu} \left \langle m_{\alpha} \Big\{
\qcq{}_{\mu}^{(\alpha)}  Q{}_{\nu}^{(\alpha)}
- \big( \qcqvec{}^{(\alpha)} \! \cdot \vec{Q}{}^{(\alpha)}
\big)
\delta_{\mu \nu}
\Big\} \right \rangle \left \langle \Big\{
\vec{I}^{-1} \qctau
\Big\}_{\nu} \right \rangle \no \\
& \qquad {} \simeq \sum_{\alpha} \left \langle\left\{
{-\pder{U}{\qcqvec^{\mathrlap{(\alpha)}\ }}\mkern8mu }
\times \vec{Q}{}^{(\alpha)} - \qfint \times  \vec{Q}{}^{(\alpha)}
\right\}_{\!\mu}\right \rangle,
\end{align}
since the variance of $\vec{Q}^{(\alpha)}$ around $\qcqvec{}^{(\alpha)}$ is expected to be small. Combining 
Eqs.~(\ref{keir}) and (\ref{resign}) leads to \eqn{tarquin}, which we now see should be a reasonable
estimator for the quasicentroid torque,
 provided the quasicentroid-constrained distribution remains compact.  

\section{\label{sec:simulations}%
Tests for liquid water and ice
}

To test the new $\vec{I}^{-1} \qctau$ estimator, we recalculated the \mbox{qTIP4P/F} infrared spectra of liquid water (300 K) and ice I$_\text{h}$ (150 K)  using the adiabatic QCMD (AQCMD) algorithm of \Refx{Trenins2019}. We also took the opportunity to improve the efficiency and stability of the algorithm, by modifying the mass scaling.

\subsection{\label{ssec:m-weighting}%
Modified mass scaling}

The AQCMD algorithm is a modification of the ACMD algorithm.\cite{Hone2004,Hone2006} The latter samples the (Cartesian) centroid-constrained ring-polymer distribution adiabatically, on the fly, by scaling the masses of the ring-polymer normal modes orthogonal to the centroid (to increase their vibrational frequencies) and thermostatting them aggressively. AQCMD applies the analogous procedure
to generate the PMF of the quasicentroid on the fly. It achieves this by mass-scaling all ring-polymer degrees of freedom (including the Cartesian centroids) and by applying quasicentroid constraints to the thermostatted ring-polymer dynamics. The resulting AQCMD algorithm
resembles the dynamics of two different systems (the ring-polymer beads and the quasicentroids) evolving in parallel. 

The original version of ACMD scaled the masses $m$ associated with the ring-polymer normal modes  orthogonal to the centroid 
($n = \pm 1, \ldots, \pm \tfrac{N-2}{2}, \tfrac{N}{2})$\footnote{Here we take $N$ to be even.} to 
\begin{equation}
\label{eq:old-scale}
m_n = \frac{m}{\kappa_n^2}, \quad 
\kappa_n = \gamma \frac{\omega_N}{\omega_n},
\end{equation}
where $\omega_n = 2 \omega_N   \sin ( \pi |n|/N) $ is the ring-polymer-spring frequency of mode $n$, $\omega_N \equiv N / \beta \hbar$, and $\gamma$ is the adiabaticity parameter. The limit $\gamma \to\infty$ corresponds to
complete adiabatic separation between the dynamics of the centroid and the non-centroid modes,
  but in practice $\gamma= 10$--$100$ is usually  sufficient numerically. One aims to keep $\gamma$ as small
   as possible in order to use the largest possible timestep. The version of AQCMD of \Refx{Trenins2019} used the 
   scaling in \eqn{eq:old-scale} for the $n\ne0$ modes, and additionally scaled the centroid ($n=0$) mass by $\kappa_0=\gamma$.

\begin{figure}[t]
\includegraphics{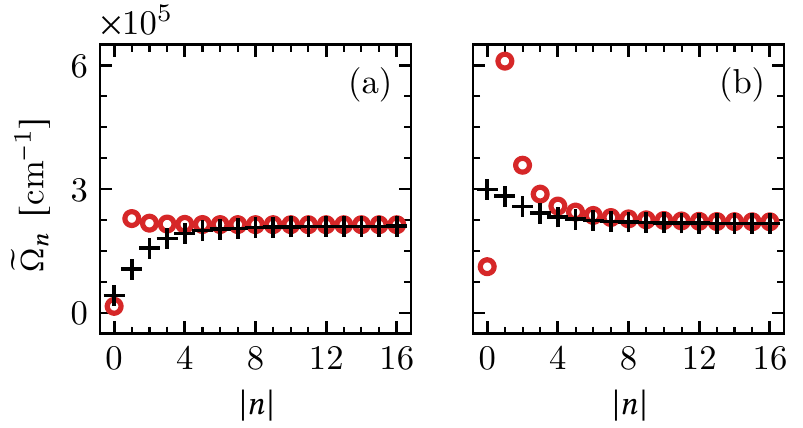}
\caption{\label{fig:mscale}
Scaled ring-polymer normal-mode vibrational frequencies for a harmonic potential with frequency
(a)~$\Omega = 500~\wns$ and (b)~$\Omega = 3500~\wns$, calculated at
a temperature $T = 300~\si{K}$ for a ring polymer of $N = 32$ beads, with $\gamma= 32$. The
red circles correspond to frequencies that have been scaled according to
the original AQCMD scheme (\eqn{eq:old-scale} for $n \neq 0$ and 
$\kappa_n = \gamma$ for $n = 0$). The black crosses correspond to
the modified scaling of \eqn{eq:new-scale} with $\Omega_\text{ref} = 2500~\wns$.
}
\end{figure}

The problem with this choice of scaling (for both ACMD and AQCMD) is that it does not scale the frequencies
 of different normal modes evenly. For a harmonic component of $V$ of frequency $\Omega$, the resulting scaled frequencies
  are
\begin{equation}
\widetilde{\Omega}_n = \kappa_n \sqrt{\omega_n^2 + \Omega^2}.
\end{equation}
(where the $\omega_n^2$ originates from the spring potential $S$).
\Fig{fig:mscale} shows that ${\Omega}_n$ has a flat distribution for $\Omega=500$~cm$^{-1}$ (roughly the libration frequency of water),
 but that it has a spike at low $n$ for  $\Omega=3500$ cm$^{-1}$ (roughly the OH stretch fequency). As a result, $\widetilde{\Omega}_n$  becomes larger than numerically necessary for low $n$, resulting in the need for a particularly small timestep.  Also, in AQCMD the (arbitrarily chosen) $\kappa_0=\gamma$ scaling of the centroid shifts the frequency of this mode significantly less than 
 that of the other modes.
 
We thus use a new mass-scaling similar to that employed in the i-PI package\cite{Kapil2019} implementation of ACMD.\footnote{This option is invoked in i-PI by setting the \texttt{style} attribute of normal-mode frequencies to
\texttt{wmax-cmd}} We take
\begin{equation}
\label{eq:new-scale}
\kappa_n = \frac{\gamma \omega_N}{\sqrt{\omega_n^2 + \Omega_\text{ref}^2}}.
\end{equation} 
for all $n$ (including $n=0$).
Setting the `reference frequency' $\Omega_\text{ref} = 2500~\wns$,\footnote{The value of $\Omega_\text{ref}$ is chosen to give a reasonably flat distribution of $\Omega_n$ over the frequency range of interest: there is no need to tune it to any characteristic frequency in the infrared spectrum. } we obtain the revised $\widetilde{\Omega}_n$
distribution shown in \fig{fig:mscale} (black crosses). Clearly the revised  $\widetilde{\Omega}_n$ is a much flatter function of $n$ for
\mbox{$\Omega = 3500~\wns$}
and can thus be expected to allow larger timesteps for a given choice of $\gamma$. 

\subsection{\label{ssec:results}%
Revised spectra for liquid water and ice}

\begin{figure}[t]
\includegraphics{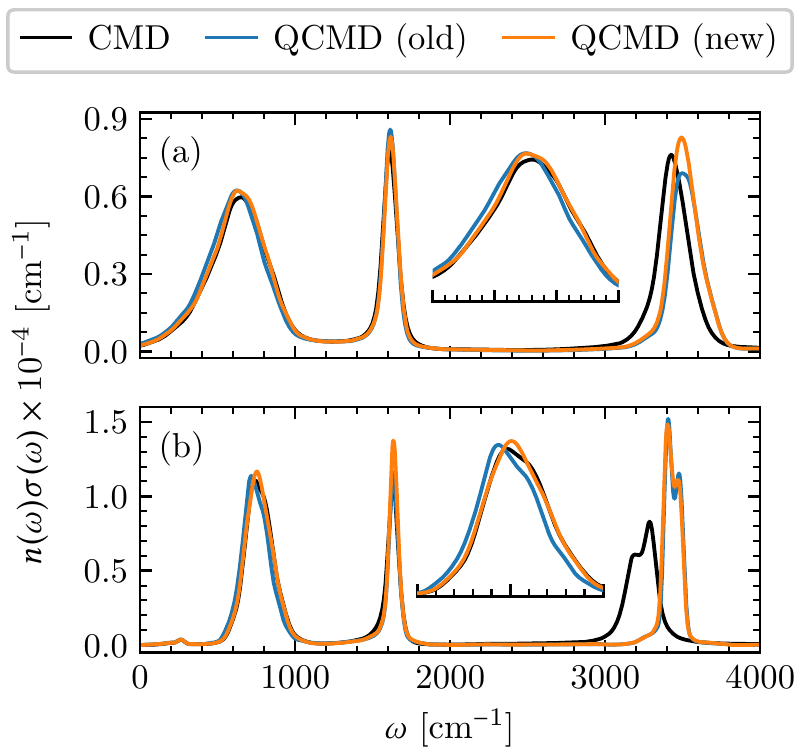}
\caption{\label{fig:spectra}
Simulated infrared absorption spectra for the \mbox{qTIP4P/F}
PES\cite{Habershon2009} for (a)~liquid water
at 300~K and (b)~ice~I\textsubscript{h} at 150 K, showing the difference made by replacing the quasicentroid torque estimator of  \Refx{Trenins2019} (old) with the new estimator of \sec{sec:torque} (new).  The insets show a magnified view of the 
libration bands, spanning the 250--1000~\wns\ and 500--1000~\wns\ regions in panels (a) and (b) respectively. The CMD and QCMD libration bands become practically identical
when the revised QCMD torques are used.
}
\end{figure}

The QCMD spectra simulated using the new $\vec{I}^{-1} \qctau$ estimator for \mbox{qTIP4P/F} liquid water at $300~\si{K}$
and ice I\textsubscript{h} at $150~\si{K}$ are shown in \fig{fig:spectra}, where they are compared with the original
QCMD simulations of \Refx{Trenins2019} and with the results of CMD.
The modified mass scaling of \eqn{eq:new-scale} was found to reduce computational cost by a factor of four with respect
to the old mass scaling of \eqn{eq:old-scale}.
The original calculations used a propagation timestep of $0.1/\gamma~\si{fs}$
with $\gamma=32$ at 300~K and $\gamma = 128$ at 150~K, whereas the 
new calculations used a timestep of $0.2/\gamma~\si{fs}$
with $\gamma=16$ and $64$  at 300 and 150~K respectively.
As well as changing the mass scaling, we also re-ordered the steps in the AQCMD propagator
 from the OBABO splitting of \Refx{Trenins2019} (where O B A refer to the
  thermostat, momentum update and position update steps of the velocity Verlet propagator) to BAOAB,\cite{Leimkuhler2012,Leimkuhler2013,Leimkuhler2016}  which was found to improve numerical stability. 

Most of the other simulation details remained the same as in \Refx{Trenins2019}.
For both water and ice I\textsubscript{h}, the simulations were initialised using
eight ring-polymer configurations that were independently pre-equilibrated 
 following standard PIMD procedure. The initial water 
  configurations were then propagated for 50~ps using the AQCMD algorithm. The ice configurations were first thermalised for 1~ps, with a local Langevin thermostat acting on the
quasicentroids; this was followed by a 5~ps production run using a global quasicentroid
Langevin thermostat;\cite{Bussi2008} the thermalisation--production cycle was then
repeated another four times. The IR absorption spectra were calculated from 
the average quasicentroid dipole-derivative time-correlation function, following Appendix~A of \Refx{Trenins2019}.

\Fig{fig:spectra} shows that use of
 the new $\vec{I}^{-1} \qctau$ estimator has eliminated the artificial 25 cm$^{-1}$ red shift from the  QCMD 
  libration bands at both temperatures, which now overlap almost exactly with CMD. 
At higher frequencies the original and revised QCMD spectra are identical
to within statistical uncertainty, except for the intensity of the
OH stretch ($\approx 3500~\wns$) at 300~K, which has increased slightly in the revised
spectrum; this brings the intensity ratio of the HOH bend and the OH stretch
more in line with those observed in classical MD and CMD calculations at the
same temperature.\cite{Benson2019}
For these reasons, we believe that the new $\vec{I}^{-1} \qctau$ estimator  gives a better approximation to the exact torque on the molecule quasicentroids than the old estimator of Ref.~\onlinecite{Trenins2019}.

\subsection{Radial distribution functions}

\begin{figure}[t]
\includegraphics{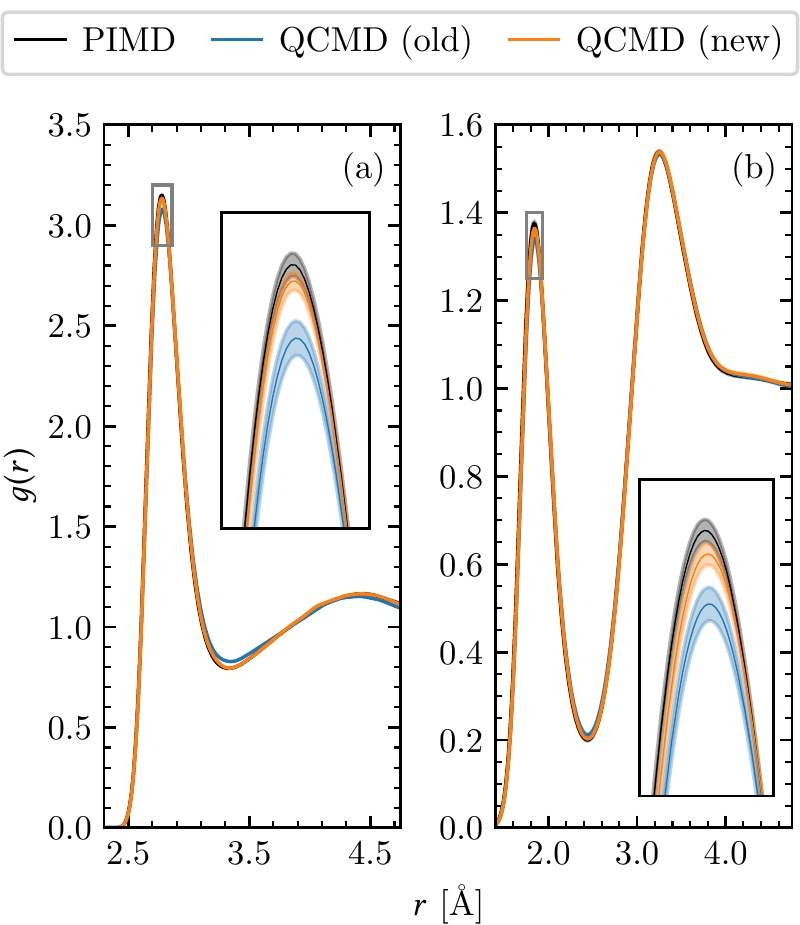}
\caption{\label{fig:rdfs} Simulated (a) oxygen-oxygen (O–O) and (b) oxygen-hydrogen (O–H) radial distribution functions (RDFs) for \mbox{qTIP4P/F} water at 300 K, calculated using the old and new quasicentroid torque estimators and standard  PIMD. The  QCMD  results  are  given  for  an  adiabatic  separation  of $\gamma = 32$ with 32 replicas. The insets show the portion of the first RDF peak highlighted by the grey box.
}
\end{figure}

To consolidate this last statement, we also extracted the (static) O--H and O--O radial distribution functions (RDFs) from the QCMD calculations at  $300~\si{K}$, and compared them with those computed using the old estimator of Ref.~\onlinecite{Trenins2019} and standard PIMD (\fig{fig:rdfs}). 
Pleasingly, the new estimator reduces the small errors in the radial distribution functions, which now follow the PIMD results very closely. 

More broadly, these RDF results show that purely static properties are sufficiently sensitive to the torque estimators
that they can be used to verify (independently of the heuristic derivation of \sec{sec:torque}) which estimator
gives a better description of the torque on the quasicentroids.  This property should be
useful when extending QCMD to treat condensed-phase systems other than pure water.

\section{\label{sec:conclusions}%
Conclusions}

The QCMD calculations of \Refx{Trenins2019}, which use the old estimator for $\qctau$, have already given very promising results, yielding infrared 
spectra for \mbox{qTIP4P/F} water and ice which eliminate most of the artefacts in the stretch region associated with CMD and (T)RPMD. 
However, it is reassuring that the one minor anomaly in these spectra, namely the 25 cm$^{-1}$ red shift of the librational
bands, is removed by using the improved estimator presented here, bringing this (essentially classical) region of the spectrum in line with the results of
 CMD, (T)RPMD and classical MD calculations.

It is also useful that we have been able to speed up the AQCMD algorithm by a factor of four using the modified `i-PI' mass scaling
of \sec{ssec:m-weighting}. However, even with this gain in efficiency,  AQCMD remains very costly:  
it is twice as expensive as adiabatic CMD for \mbox{qTIP4P/F} liquid water at 300 K and eight times for ice at 150 K. Another drawback of QCMD is  that it requires the internal quasicentroid coordinates to be tailored to each system and is thus tricky to generalise. Nevertheless, recent developments,
 especially of the fast QCMD method (f-QCMD),\cite{Fletcher2021} suggest that QCMD is likely to become a powerful method for computing  infrared spectra in the condensed phase, provided the dynamics do not involve strongly anharmonic `floppy' motion 
 such as in protonated water clusters.\cite{Yu2019} We hope that the calculations presented
  here will serve as useful benchmarks for these and related\cite{Venkat2022} future developments.

 \begin{acknowledgments}
 G.T. acknowledges support from the Cambridge Philosophical Society and St Catharine’s College, University of Cambridge. C.H. acknowledges support from the EPSRC Centre for Doctoral Training in Computational Methods for Materials Science (Grant No. EP/L015552/1).
 \end{acknowledgments}

\bibliography{references}

\begin{thebibliography}{50}%
\makeatletter
\providecommand \@ifxundefined [1]{%
 \@ifx{#1\undefined}
}%
\providecommand \@ifnum [1]{%
 \ifnum #1\expandafter \@firstoftwo
 \else \expandafter \@secondoftwo
 \fi
}%
\providecommand \@ifx [1]{%
 \ifx #1\expandafter \@firstoftwo
 \else \expandafter \@secondoftwo
 \fi
}%
\providecommand \natexlab [1]{#1}%
\providecommand \enquote  [1]{``#1''}%
\providecommand \bibnamefont  [1]{#1}%
\providecommand \bibfnamefont [1]{#1}%
\providecommand \citenamefont [1]{#1}%
\providecommand \href@noop [0]{\@secondoftwo}%
\providecommand \href [0]{\begingroup \@sanitize@url \@href}%
\providecommand \@href[1]{\@@startlink{#1}\@@href}%
\providecommand \@@href[1]{\endgroup#1\@@endlink}%
\providecommand \@sanitize@url [0]{\catcode `\\12\catcode `\$12\catcode
  `\&12\catcode `\#12\catcode `\^12\catcode `\_12\catcode `\%12\relax}%
\providecommand \@@startlink[1]{}%
\providecommand \@@endlink[0]{}%
\providecommand \url  [0]{\begingroup\@sanitize@url \@url }%
\providecommand \@url [1]{\endgroup\@href {#1}{\urlprefix }}%
\providecommand \urlprefix  [0]{URL }%
\providecommand \Eprint [0]{\href }%
\providecommand \doibase [0]{https://doi.org/}%
\providecommand \selectlanguage [0]{\@gobble}%
\providecommand \bibinfo  [0]{\@secondoftwo}%
\providecommand \bibfield  [0]{\@secondoftwo}%
\providecommand \translation [1]{[#1]}%
\providecommand \BibitemOpen [0]{}%
\providecommand \bibitemStop [0]{}%
\providecommand \bibitemNoStop [0]{.\EOS\space}%
\providecommand \EOS [0]{\spacefactor3000\relax}%
\providecommand \BibitemShut  [1]{\csname bibitem#1\endcsname}%
\let\auto@bib@innerbib\@empty
\bibitem [{\citenamefont {Trenins}, \citenamefont {Willatt},\ and\
  \citenamefont {Althorpe}(2019)}]{Trenins2019}%
  \BibitemOpen
  \bibfield  {author} {\bibinfo {author} {\bibfnamefont {G.}~\bibnamefont
  {Trenins}}, \bibinfo {author} {\bibfnamefont {M.~J.}\ \bibnamefont
  {Willatt}},\ and\ \bibinfo {author} {\bibfnamefont {S.~C.}\ \bibnamefont
  {Althorpe}},\ }\bibfield  {title} {\enquote {\bibinfo {title} {Path-integral
  dynamics of water using curvilinear centroids},}\ }\href
  {https://doi.org/10.1063/1.5100587} {\bibfield  {journal} {\bibinfo
  {journal} {J. Chem. Phys.}\ }\textbf {\bibinfo {volume} {151}},\ \bibinfo
  {pages} {054109} (\bibinfo {year} {2019})}\BibitemShut {NoStop}%
\bibitem [{\citenamefont {Benson}, \citenamefont {Trenins},\ and\ \citenamefont
  {Althorpe}(2020)}]{Benson2019}%
  \BibitemOpen
  \bibfield  {author} {\bibinfo {author} {\bibfnamefont {R.~L.}\ \bibnamefont
  {Benson}}, \bibinfo {author} {\bibfnamefont {G.}~\bibnamefont {Trenins}},\
  and\ \bibinfo {author} {\bibfnamefont {S.~C.}\ \bibnamefont {Althorpe}},\
  }\bibfield  {title} {\enquote {\bibinfo {title} {Which quantum
  statistics--classical dynamics method is best for water?}}\ }\href
  {https://doi.org/10.1039/C9FD00077A} {\bibfield  {journal} {\bibinfo
  {journal} {Faraday Discuss.}\ }\textbf {\bibinfo {volume} {221}},\ \bibinfo
  {pages} {350} (\bibinfo {year} {2020})}\BibitemShut {NoStop}%
\bibitem [{\citenamefont {Haggard}\ \emph {et~al.}(2021)\citenamefont
  {Haggard}, \citenamefont {Sadhasivam}, \citenamefont {Trenins},\ and\
  \citenamefont {Althorpe}}]{Haggard2021}%
  \BibitemOpen
  \bibfield  {author} {\bibinfo {author} {\bibfnamefont {C.}~\bibnamefont
  {Haggard}}, \bibinfo {author} {\bibfnamefont {V.~G.}\ \bibnamefont
  {Sadhasivam}}, \bibinfo {author} {\bibfnamefont {G.}~\bibnamefont
  {Trenins}},\ and\ \bibinfo {author} {\bibfnamefont {S.~C.}\ \bibnamefont
  {Althorpe}},\ }\bibfield  {title} {\enquote {\bibinfo {title} {Testing the
  quasicentroid molecular dynamics method on gas-phase ammonia},}\ }\href
  {https://doi.org/10.1063/5.0068250} {\bibfield  {journal} {\bibinfo
  {journal} {J. Chem. Phys.}\ }\textbf {\bibinfo {volume} {155}},\ \bibinfo
  {pages} {174120} (\bibinfo {year} {2021})}\BibitemShut {NoStop}%
\bibitem [{\citenamefont {Fletcher}\ \emph {et~al.}(2021)\citenamefont
  {Fletcher}, \citenamefont {Zhu}, \citenamefont {Lawrence},\ and\
  \citenamefont {Manolopoulos}}]{Fletcher2021}%
  \BibitemOpen
  \bibfield  {author} {\bibinfo {author} {\bibfnamefont {T.}~\bibnamefont
  {Fletcher}}, \bibinfo {author} {\bibfnamefont {A.}~\bibnamefont {Zhu}},
  \bibinfo {author} {\bibfnamefont {J.~E.}\ \bibnamefont {Lawrence}},\ and\
  \bibinfo {author} {\bibfnamefont {D.~E.}\ \bibnamefont {Manolopoulos}},\
  }\bibfield  {title} {\enquote {\bibinfo {title} {Fast quasi-centroid
  molecular dynamics},}\ }\href {https://doi.org/10.1063/5.0076704} {\bibfield
  {journal} {\bibinfo  {journal} {J. Chem. Phys.}\ }\textbf {\bibinfo {volume}
  {155}},\ \bibinfo {pages} {231101} (\bibinfo {year} {2021})}\BibitemShut
  {NoStop}%
\bibitem [{\citenamefont {Habershon}, \citenamefont {Fanourgakis},\ and\
  \citenamefont {Manolopoulos}(2008)}]{Habershon2008}%
  \BibitemOpen
  \bibfield  {author} {\bibinfo {author} {\bibfnamefont {S.}~\bibnamefont
  {Habershon}}, \bibinfo {author} {\bibfnamefont {G.~S.}\ \bibnamefont
  {Fanourgakis}},\ and\ \bibinfo {author} {\bibfnamefont {D.~E.}\ \bibnamefont
  {Manolopoulos}},\ }\bibfield  {title} {\enquote {\bibinfo {title} {Comparison
  of path integral molecular dynamics methods for the infrared absorption
  spectrum of liquid water},}\ }\href {https://doi.org/10.1063/1.2968555}
  {\bibfield  {journal} {\bibinfo  {journal} {J. Chem. Phys.}\ }\textbf
  {\bibinfo {volume} {129}},\ \bibinfo {pages} {074501} (\bibinfo {year}
  {2008})}\BibitemShut {NoStop}%
\bibitem [{\citenamefont {Witt}\ \emph {et~al.}(2009)\citenamefont {Witt},
  \citenamefont {Ivanov}, \citenamefont {Shiga}, \citenamefont {Forbert},\ and\
  \citenamefont {Marx}}]{Witt2009}%
  \BibitemOpen
  \bibfield  {author} {\bibinfo {author} {\bibfnamefont {A.}~\bibnamefont
  {Witt}}, \bibinfo {author} {\bibfnamefont {S.~D.}\ \bibnamefont {Ivanov}},
  \bibinfo {author} {\bibfnamefont {M.}~\bibnamefont {Shiga}}, \bibinfo
  {author} {\bibfnamefont {H.}~\bibnamefont {Forbert}},\ and\ \bibinfo {author}
  {\bibfnamefont {D.}~\bibnamefont {Marx}},\ }\bibfield  {title} {\enquote
  {\bibinfo {title} {On the applicability of centroid and ring polymer path
  integral molecular dynamics for vibrational spectroscopy},}\ }\href
  {https://doi.org/10.1063/1.3125009} {\bibfield  {journal} {\bibinfo
  {journal} {J. Chem. Phys.}\ }\textbf {\bibinfo {volume} {130}},\ \bibinfo
  {pages} {194510} (\bibinfo {year} {2009})}\BibitemShut {NoStop}%
\bibitem [{\citenamefont {Ivanov}\ \emph {et~al.}(2010)\citenamefont {Ivanov},
  \citenamefont {Witt}, \citenamefont {Shiga},\ and\ \citenamefont
  {Marx}}]{Ivanov2010}%
  \BibitemOpen
  \bibfield  {author} {\bibinfo {author} {\bibfnamefont {S.~D.}\ \bibnamefont
  {Ivanov}}, \bibinfo {author} {\bibfnamefont {A.}~\bibnamefont {Witt}},
  \bibinfo {author} {\bibfnamefont {M.}~\bibnamefont {Shiga}},\ and\ \bibinfo
  {author} {\bibfnamefont {D.}~\bibnamefont {Marx}},\ }\bibfield  {title}
  {\enquote {\bibinfo {title} {Communications: {O}n artificial frequency shifts
  in infrared spectra obtained from centroid molecular dynamics: {Q}uantum
  liquid water},}\ }\href {https://doi.org/10.1063/1.3290958} {\bibfield
  {journal} {\bibinfo  {journal} {J. Chem. Phys.}\ }\textbf {\bibinfo {volume}
  {132}},\ \bibinfo {pages} {031101} (\bibinfo {year} {2010})}\BibitemShut
  {NoStop}%
\bibitem [{\citenamefont {Rossi}, \citenamefont {Ceriotti},\ and\ \citenamefont
  {Manolopoulos}(2014)}]{Rossi2014}%
  \BibitemOpen
  \bibfield  {author} {\bibinfo {author} {\bibfnamefont {M.}~\bibnamefont
  {Rossi}}, \bibinfo {author} {\bibfnamefont {M.}~\bibnamefont {Ceriotti}},\
  and\ \bibinfo {author} {\bibfnamefont {D.~E.}\ \bibnamefont {Manolopoulos}},\
  }\bibfield  {title} {\enquote {\bibinfo {title} {How to remove the spurious
  resonances from ring polymer molecular dynamics},}\ }\href
  {https://doi.org/10.1063/1.4883861} {\bibfield  {journal} {\bibinfo
  {journal} {J. Chem. Phys.}\ }\textbf {\bibinfo {volume} {140}},\ \bibinfo
  {pages} {234116} (\bibinfo {year} {2014})}\BibitemShut {NoStop}%
\bibitem [{\citenamefont {Trenins}\ and\ \citenamefont
  {Althorpe}(2018)}]{Trenins2018}%
  \BibitemOpen
  \bibfield  {author} {\bibinfo {author} {\bibfnamefont {G.}~\bibnamefont
  {Trenins}}\ and\ \bibinfo {author} {\bibfnamefont {S.~C.}\ \bibnamefont
  {Althorpe}},\ }\bibfield  {title} {\enquote {\bibinfo {title} {Mean-field
  {M}atsubara dynamics: Analysis of path-integral curvature effects in
  rovibrational spectra},}\ }\href {https://doi.org/10.1063/1.5038616}
  {\bibfield  {journal} {\bibinfo  {journal} {J. Chem. Phys.}\ }\textbf
  {\bibinfo {volume} {149}},\ \bibinfo {pages} {014102} (\bibinfo {year}
  {2018})}\BibitemShut {NoStop}%
\bibitem [{\citenamefont {Rossi}, \citenamefont {Kapil},\ and\ \citenamefont
  {Ceriotti}(2018)}]{Rossi2018}%
  \BibitemOpen
  \bibfield  {author} {\bibinfo {author} {\bibfnamefont {M.}~\bibnamefont
  {Rossi}}, \bibinfo {author} {\bibfnamefont {V.}~\bibnamefont {Kapil}},\ and\
  \bibinfo {author} {\bibfnamefont {M.}~\bibnamefont {Ceriotti}},\ }\bibfield
  {title} {\enquote {\bibinfo {title} {Fine tuning classical and quantum
  molecular dynamics using a generalized {L}angevin equation},}\ }\href
  {https://doi.org/10.1063/1.4990536} {\bibfield  {journal} {\bibinfo
  {journal} {J. Chem. Phys.}\ }\textbf {\bibinfo {volume} {148}},\ \bibinfo
  {pages} {102301} (\bibinfo {year} {2018})}\BibitemShut {NoStop}%
\bibitem [{\citenamefont {Cao}\ and\ \citenamefont {Voth}(1994)}]{Cao1994b}%
  \BibitemOpen
  \bibfield  {author} {\bibinfo {author} {\bibfnamefont {J.}~\bibnamefont
  {Cao}}\ and\ \bibinfo {author} {\bibfnamefont {G.~A.}\ \bibnamefont {Voth}},\
  }\bibfield  {title} {\enquote {\bibinfo {title} {The formulation of quantum
  statistical mechanics based on the {F}eynman path centroid density. {IV}.
  {A}lgorithms for centroid molecular dynamics},}\ }\href
  {https://doi.org/10.1063/1.468399} {\bibfield  {journal} {\bibinfo  {journal}
  {J. Chem. Phys.}\ }\textbf {\bibinfo {volume} {101}},\ \bibinfo {pages}
  {6168} (\bibinfo {year} {1994})}\BibitemShut {NoStop}%
\bibitem [{\citenamefont {Hone}\ and\ \citenamefont {Voth}(2004)}]{Hone2004}%
  \BibitemOpen
  \bibfield  {author} {\bibinfo {author} {\bibfnamefont {T.~D.}\ \bibnamefont
  {Hone}}\ and\ \bibinfo {author} {\bibfnamefont {G.~A.}\ \bibnamefont
  {Voth}},\ }\bibfield  {title} {\enquote {\bibinfo {title} {A centroid
  molecular dynamics study of liquid para-hydrogen and ortho-deuterium},}\
  }\href {https://doi.org/10.1063/1.1780951} {\bibfield  {journal} {\bibinfo
  {journal} {J. Chem. Phys.}\ }\textbf {\bibinfo {volume} {121}},\ \bibinfo
  {pages} {6412} (\bibinfo {year} {2004})}\BibitemShut {NoStop}%
\bibitem [{\citenamefont {Hone}, \citenamefont {Rossky},\ and\ \citenamefont
  {Voth}(2006)}]{Hone2006}%
  \BibitemOpen
  \bibfield  {author} {\bibinfo {author} {\bibfnamefont {T.~D.}\ \bibnamefont
  {Hone}}, \bibinfo {author} {\bibfnamefont {P.~J.}\ \bibnamefont {Rossky}},\
  and\ \bibinfo {author} {\bibfnamefont {G.~A.}\ \bibnamefont {Voth}},\
  }\bibfield  {title} {\enquote {\bibinfo {title} {A comparative study of
  imaginary time path integral based methods for quantum dynamics},}\ }\href
  {https://doi.org/10.1063/1.2186636} {\bibfield  {journal} {\bibinfo
  {journal} {J. Chem. Phys.}\ }\textbf {\bibinfo {volume} {124}},\ \bibinfo
  {pages} {154103} (\bibinfo {year} {2006})}\BibitemShut {NoStop}%
\bibitem [{\citenamefont {Craig}\ and\ \citenamefont
  {Manolopoulos}(2004)}]{Craig2004}%
  \BibitemOpen
  \bibfield  {author} {\bibinfo {author} {\bibfnamefont {I.~R.}\ \bibnamefont
  {Craig}}\ and\ \bibinfo {author} {\bibfnamefont {D.~E.}\ \bibnamefont
  {Manolopoulos}},\ }\bibfield  {title} {\enquote {\bibinfo {title} {Quantum
  statistics and classical mechanics: Real time correlation functions from ring
  polymer molecular dynamics},}\ }\href {https://doi.org/10.1063/1.1777575}
  {\bibfield  {journal} {\bibinfo  {journal} {J. Chem. Phys.}\ }\textbf
  {\bibinfo {volume} {121}},\ \bibinfo {pages} {3368} (\bibinfo {year}
  {2004})}\BibitemShut {NoStop}%
\bibitem [{\citenamefont {Habershon}\ \emph {et~al.}(2012)\citenamefont
  {Habershon}, \citenamefont {Manolopoulos}, \citenamefont {Markland},\ and\
  \citenamefont {Miller~III}}]{Habershon2012}%
  \BibitemOpen
  \bibfield  {author} {\bibinfo {author} {\bibfnamefont {S.}~\bibnamefont
  {Habershon}}, \bibinfo {author} {\bibfnamefont {D.~E.}\ \bibnamefont
  {Manolopoulos}}, \bibinfo {author} {\bibfnamefont {T.~E.}\ \bibnamefont
  {Markland}},\ and\ \bibinfo {author} {\bibfnamefont {T.~F.}\ \bibnamefont
  {Miller~III}},\ }\bibfield  {title} {\enquote {\bibinfo {title} {Ring-polymer
  molecular dynamics: {Q}uantum effects in chemical dynamics from classical
  trajectories in an extended phase space},}\ }\href
  {http://www.annualreviews.org/doi/abs/10.1146/annurev-physchem-040412-110122}
  {\bibfield  {journal} {\bibinfo  {journal} {Annu. Rev. Phys. Chem.}\ }\textbf
  {\bibinfo {volume} {64}},\ \bibinfo {pages} {387} (\bibinfo {year}
  {2012})}\BibitemShut {NoStop}%
\bibitem [{\citenamefont {Miller~III}\ and\ \citenamefont
  {Manolopoulos}(2005)}]{MillerThomasF2005b}%
  \BibitemOpen
  \bibfield  {author} {\bibinfo {author} {\bibfnamefont {T.~F.}\ \bibnamefont
  {Miller~III}}\ and\ \bibinfo {author} {\bibfnamefont {D.~E.}\ \bibnamefont
  {Manolopoulos}},\ }\bibfield  {title} {\enquote {\bibinfo {title} {Quantum
  diffusion in liquid para-hydrogen from ring-polymer molecular dynamics},}\
  }\href {papers2://publication/doi/10.1063/1.1893956} {\bibfield  {journal}
  {\bibinfo  {journal} {J. Chem. Phys.}\ }\textbf {\bibinfo {volume} {122}},\
  \bibinfo {pages} {184503} (\bibinfo {year} {2005})}\BibitemShut {NoStop}%
\bibitem [{\citenamefont {Habershon}, \citenamefont {Markland},\ and\
  \citenamefont {Manolopoulos}(2009)}]{Habershon2009}%
  \BibitemOpen
  \bibfield  {author} {\bibinfo {author} {\bibfnamefont {S.}~\bibnamefont
  {Habershon}}, \bibinfo {author} {\bibfnamefont {T.~E.}\ \bibnamefont
  {Markland}},\ and\ \bibinfo {author} {\bibfnamefont {D.~E.}\ \bibnamefont
  {Manolopoulos}},\ }\bibfield  {title} {\enquote {\bibinfo {title} {Competing
  quantum effects in the dynamics of a flexible water model},}\ }\href
  {https://doi.org/10.1063/1.3167790} {\bibfield  {journal} {\bibinfo
  {journal} {J. Chem. Phys.}\ }\textbf {\bibinfo {volume} {131}},\ \bibinfo
  {pages} {024501} (\bibinfo {year} {2009})}\BibitemShut {NoStop}%
\bibitem [{Note1()}]{Note1}%
  \BibitemOpen
  \bibinfo {note} {To date, f-QCMD has been applied only in the gas-phase, but
  condensed-phase applications are likely soon.}\BibitemShut {Stop}%
\bibitem [{\citenamefont {Pl{\'{e}}}\ \emph {et~al.}(2021)\citenamefont
  {Pl{\'{e}}}, \citenamefont {Huppert}, \citenamefont {Finocchi}, \citenamefont
  {Depondt},\ and\ \citenamefont {Bonella}}]{Ple2021}%
  \BibitemOpen
  \bibfield  {author} {\bibinfo {author} {\bibfnamefont {T.}~\bibnamefont
  {Pl{\'{e}}}}, \bibinfo {author} {\bibfnamefont {S.}~\bibnamefont {Huppert}},
  \bibinfo {author} {\bibfnamefont {F.}~\bibnamefont {Finocchi}}, \bibinfo
  {author} {\bibfnamefont {P.}~\bibnamefont {Depondt}},\ and\ \bibinfo {author}
  {\bibfnamefont {S.}~\bibnamefont {Bonella}},\ }\bibfield  {title} {\enquote
  {\bibinfo {title} {Anharmonic spectral features via trajectory-based quantum
  dynamics: A perturbative analysis of the interplay between dynamics and
  sampling},}\ }\href {https://doi.org/10.1063/5.0056824} {\bibfield  {journal}
  {\bibinfo  {journal} {J. Chem. Phys.}\ }\textbf {\bibinfo {volume} {155}},\
  \bibinfo {pages} {104108} (\bibinfo {year} {2021})}\BibitemShut {NoStop}%
\bibitem [{\citenamefont {Benson}\ and\ \citenamefont
  {Althorpe}(2021)}]{Benson2021}%
  \BibitemOpen
  \bibfield  {author} {\bibinfo {author} {\bibfnamefont {R.~L.}\ \bibnamefont
  {Benson}}\ and\ \bibinfo {author} {\bibfnamefont {S.~C.}\ \bibnamefont
  {Althorpe}},\ }\bibfield  {title} {\enquote {\bibinfo {title} {On the
  ``{Matsubara} heating'' of overtone intensities and {F}ermi splittings},}\
  }\href {https://doi.org/10.1063/5.0056829} {\bibfield  {journal} {\bibinfo
  {journal} {J. Chem. Phys.}\ }\textbf {\bibinfo {volume} {155}},\ \bibinfo
  {pages} {104107} (\bibinfo {year} {2021})}\BibitemShut {NoStop}%
\bibitem [{\citenamefont {Hele}\ \emph
  {et~al.}(2015{\natexlab{a}})\citenamefont {Hele}, \citenamefont {Willatt},
  \citenamefont {Muolo},\ and\ \citenamefont {Althorpe}}]{Hele2015}%
  \BibitemOpen
  \bibfield  {author} {\bibinfo {author} {\bibfnamefont {T.~J.~H.}\
  \bibnamefont {Hele}}, \bibinfo {author} {\bibfnamefont {M.~J.}\ \bibnamefont
  {Willatt}}, \bibinfo {author} {\bibfnamefont {A.}~\bibnamefont {Muolo}},\
  and\ \bibinfo {author} {\bibfnamefont {S.~C.}\ \bibnamefont {Althorpe}},\
  }\bibfield  {title} {\enquote {\bibinfo {title} {Boltzmann-conserving
  classical dynamics in quantum time-correlation functions: ``{Matsubara}
  dynamics''},}\ }\href {https://doi.org/10.1063/1.4916311} {\bibfield
  {journal} {\bibinfo  {journal} {J. Chem. Phys.}\ }\textbf {\bibinfo {volume}
  {142}},\ \bibinfo {pages} {134103} (\bibinfo {year}
  {2015}{\natexlab{a}})}\BibitemShut {NoStop}%
\bibitem [{\citenamefont {Hele}\ \emph
  {et~al.}(2015{\natexlab{b}})\citenamefont {Hele}, \citenamefont {Willatt},
  \citenamefont {Muolo},\ and\ \citenamefont {Althorpe}}]{Hele2015a}%
  \BibitemOpen
  \bibfield  {author} {\bibinfo {author} {\bibfnamefont {T.~J.~H.}\
  \bibnamefont {Hele}}, \bibinfo {author} {\bibfnamefont {M.~J.}\ \bibnamefont
  {Willatt}}, \bibinfo {author} {\bibfnamefont {A.}~\bibnamefont {Muolo}},\
  and\ \bibinfo {author} {\bibfnamefont {S.~C.}\ \bibnamefont {Althorpe}},\
  }\bibfield  {title} {\enquote {\bibinfo {title} {Communication: {R}elation of
  centroid molecular dynamics and ring-polymer molecular dynamics to exact
  quantum dynamics},}\ }\href {https://doi.org/10.1063/1.4921234} {\bibfield
  {journal} {\bibinfo  {journal} {J. Chem. Phys.}\ }\textbf {\bibinfo {volume}
  {142}},\ \bibinfo {pages} {191101} (\bibinfo {year}
  {2015}{\natexlab{b}})}\BibitemShut {NoStop}%
\bibitem [{\citenamefont {Jung}, \citenamefont {Videla},\ and\ \citenamefont
  {Batista}(2019)}]{Jung2019}%
  \BibitemOpen
  \bibfield  {author} {\bibinfo {author} {\bibfnamefont {K.~A.}\ \bibnamefont
  {Jung}}, \bibinfo {author} {\bibfnamefont {P.~E.}\ \bibnamefont {Videla}},\
  and\ \bibinfo {author} {\bibfnamefont {V.~S.}\ \bibnamefont {Batista}},\
  }\bibfield  {title} {\enquote {\bibinfo {title} {Multi-time formulation of
  {Matsubara} dynamics},}\ }\href {https://doi.org/10.1063/1.5110427}
  {\bibfield  {journal} {\bibinfo  {journal} {J. Chem. Phys.}\ }\textbf
  {\bibinfo {volume} {151}},\ \bibinfo {pages} {034108} (\bibinfo {year}
  {2019})}\BibitemShut {NoStop}%
\bibitem [{\citenamefont {Althorpe}(2021)}]{Althorpe2021}%
  \BibitemOpen
  \bibfield  {author} {\bibinfo {author} {\bibfnamefont {S.~C.}\ \bibnamefont
  {Althorpe}},\ }\bibfield  {title} {\enquote {\bibinfo {title} {Path-integral
  approximations to quantum dynamics},}\ }\href@noop {} {\bibfield  {journal}
  {\bibinfo  {journal} {Eur. Phys. J. B}\ }\textbf {\bibinfo {volume} {94}},\
  \bibinfo {pages} {155} (\bibinfo {year} {2021})}\BibitemShut {NoStop}%
\bibitem [{Note2()}]{Note2}%
  \BibitemOpen
  \bibinfo {note} {For gas-phase water and ammonia the intensities of the
  overtone and combination bands can be corrected using harmonic perturbation
  theory;\cite {Ple2021,Benson2021} for methane this approach has proved less
  successful.\cite {Fletcher2021}}\BibitemShut {NoStop}%
\bibitem [{Note3()}]{Note3}%
  \BibitemOpen
  \bibinfo {note} {QCMD calculations using more realistic water potentials such
  as the MB-pol surface of Ref.~\protect \rev@citealpnum {Babin2014} or the DFT
  scheme of Ref.~\protect \rev@citealpnum {Marsalek2017} have not yet been
  reported.}\BibitemShut {Stop}%
\bibitem [{\citenamefont {Rossi}\ \emph {et~al.}(2014)\citenamefont {Rossi},
  \citenamefont {Liu}, \citenamefont {Paesani}, \citenamefont {Bowman},\ and\
  \citenamefont {Ceriotti}}]{Rossi2014a}%
  \BibitemOpen
  \bibfield  {author} {\bibinfo {author} {\bibfnamefont {M.}~\bibnamefont
  {Rossi}}, \bibinfo {author} {\bibfnamefont {H.}~\bibnamefont {Liu}}, \bibinfo
  {author} {\bibfnamefont {F.}~\bibnamefont {Paesani}}, \bibinfo {author}
  {\bibfnamefont {J.}~\bibnamefont {Bowman}},\ and\ \bibinfo {author}
  {\bibfnamefont {M.}~\bibnamefont {Ceriotti}},\ }\bibfield  {title} {\enquote
  {\bibinfo {title} {Communication: {O}n the consistency of approximate quantum
  dynamics simulation methods for vibrational spectra in the condensed
  phase},}\ }\href {https://doi.org/10.1063/1.4901214} {\bibfield  {journal}
  {\bibinfo  {journal} {J. Chem. Phys.}\ }\textbf {\bibinfo {volume} {141}},\
  \bibinfo {pages} {181101} (\bibinfo {year} {2014})}\BibitemShut {NoStop}%
\bibitem [{\citenamefont {Liu}, \citenamefont {Wang},\ and\ \citenamefont
  {Bowman}(2015)}]{Liu2015}%
  \BibitemOpen
  \bibfield  {author} {\bibinfo {author} {\bibfnamefont {H.}~\bibnamefont
  {Liu}}, \bibinfo {author} {\bibfnamefont {Y.}~\bibnamefont {Wang}},\ and\
  \bibinfo {author} {\bibfnamefont {J.~M.}\ \bibnamefont {Bowman}},\ }\bibfield
   {title} {\enquote {\bibinfo {title} {Transferable ab initio dipole moment
  for water: Three applications to bulk water},}\ }\href
  {http://pubs.acs.org/doi/10.1021/acs.jpcb.5b09213} {\bibfield  {journal}
  {\bibinfo  {journal} {J. Phys. Chem. B}\ }\textbf {\bibinfo {volume} {120}},\
  \bibinfo {pages} {1735} (\bibinfo {year} {2015})}\BibitemShut {NoStop}%
\bibitem [{\citenamefont {Liu}\ \emph {et~al.}(2011)\citenamefont {Liu},
  \citenamefont {Miller}, \citenamefont {Fanourgakis}, \citenamefont
  {Xantheas}, \citenamefont {Imoto},\ and\ \citenamefont {Saito}}]{Liu2011a}%
  \BibitemOpen
  \bibfield  {author} {\bibinfo {author} {\bibfnamefont {J.}~\bibnamefont
  {Liu}}, \bibinfo {author} {\bibfnamefont {W.~H.}\ \bibnamefont {Miller}},
  \bibinfo {author} {\bibfnamefont {G.~S.}\ \bibnamefont {Fanourgakis}},
  \bibinfo {author} {\bibfnamefont {S.~S.}\ \bibnamefont {Xantheas}}, \bibinfo
  {author} {\bibfnamefont {S.}~\bibnamefont {Imoto}},\ and\ \bibinfo {author}
  {\bibfnamefont {S.}~\bibnamefont {Saito}},\ }\bibfield  {title} {\enquote
  {\bibinfo {title} {Insights in quantum dynamical effects in the infrared
  spectroscopy of liquid water from a semiclassical study with an ab
  initio-based flexible and polarizable force field},}\ }\href
  {http://scitation.aip.org/content/aip/journal/jcp/135/24/10.1063/1.3670960}
  {\bibfield  {journal} {\bibinfo  {journal} {J. Chem. Phys.}\ }\textbf
  {\bibinfo {volume} {135}},\ \bibinfo {pages} {244503} (\bibinfo {year}
  {2011})}\BibitemShut {NoStop}%
\bibitem [{\citenamefont {Chandler}\ and\ \citenamefont
  {Wolynes}(1981)}]{Chandler1981}%
  \BibitemOpen
  \bibfield  {author} {\bibinfo {author} {\bibfnamefont {D.}~\bibnamefont
  {Chandler}}\ and\ \bibinfo {author} {\bibfnamefont {P.~G.}\ \bibnamefont
  {Wolynes}},\ }\bibfield  {title} {\enquote {\bibinfo {title} {Exploiting the
  isomorphism between quantum theory and classical statistical mechanics of
  polyatomic fluids},}\ }\href
  {papers2://publication/uuid/26DAD931-0646-4540-BECC-BAE86066A9CC} {\bibfield
  {journal} {\bibinfo  {journal} {J. Chem. Phys.}\ }\textbf {\bibinfo {volume}
  {74}},\ \bibinfo {pages} {4078} (\bibinfo {year} {1981})}\BibitemShut
  {NoStop}%
\bibitem [{\citenamefont {Parrinello}\ and\ \citenamefont
  {Rahman}(1984)}]{Parrinello1984}%
  \BibitemOpen
  \bibfield  {author} {\bibinfo {author} {\bibfnamefont {M.}~\bibnamefont
  {Parrinello}}\ and\ \bibinfo {author} {\bibfnamefont {A.}~\bibnamefont
  {Rahman}},\ }\bibfield  {title} {\enquote {\bibinfo {title} {Study of an {F}
  center in molten {KCl}},}\ }\href
  {http://scitation.aip.org/content/aip/journal/jcp/80/2/10.1063/1.446740}
  {\bibfield  {journal} {\bibinfo  {journal} {J. Chem. Phys.}\ }\textbf
  {\bibinfo {volume} {80}},\ \bibinfo {pages} {860} (\bibinfo {year}
  {1984})}\BibitemShut {NoStop}%
\bibitem [{Note4()}]{Note4}%
  \BibitemOpen
  \bibinfo {note} {This does not give back the Cartesian centroids owing to the
  non-linearity of the Cartesian to bond-angle coordinate
  transformation.}\BibitemShut {Stop}%
\bibitem [{\citenamefont {Allen}\ and\ \citenamefont
  {Tildesley}(2017)}]{AllenBook}%
  \BibitemOpen
  \bibfield  {author} {\bibinfo {author} {\bibfnamefont {M.}~\bibnamefont
  {Allen}}\ and\ \bibinfo {author} {\bibfnamefont {D.}~\bibnamefont
  {Tildesley}},\ }\href {https://books.google.ch/books?id=WFExDwAAQBAJ} {\emph
  {\bibinfo {title} {Computer Simulation of Liquids}}}\ (\bibinfo  {publisher}
  {Oxford University Press},\ \bibinfo {address} {Oxford},\ \bibinfo {year}
  {2017})\BibitemShut {NoStop}%
\bibitem [{\citenamefont {Eckart}(1935)}]{Eckart1935}%
  \BibitemOpen
  \bibfield  {author} {\bibinfo {author} {\bibfnamefont {C.}~\bibnamefont
  {Eckart}},\ }\bibfield  {title} {\enquote {\bibinfo {title} {Some studies
  concerning rotating axes and polyatomic molecules},}\ }\href
  {https://doi.org/10.1103/PhysRev.47.552} {\bibfield  {journal} {\bibinfo
  {journal} {Phys. Rev.}\ }\textbf {\bibinfo {volume} {47}},\ \bibinfo {pages}
  {552} (\bibinfo {year} {1935})}\BibitemShut {NoStop}%
\bibitem [{\citenamefont {Wilson}, \citenamefont {Decius},\ and\ \citenamefont
  {Cross}(1980)}]{WilsonBook}%
  \BibitemOpen
  \bibfield  {author} {\bibinfo {author} {\bibfnamefont {E.}~\bibnamefont
  {Wilson}}, \bibinfo {author} {\bibfnamefont {J.}~\bibnamefont {Decius}},\
  and\ \bibinfo {author} {\bibfnamefont {P.}~\bibnamefont {Cross}},\ }\href
  {https://books.google.ch/books?id=yKVPDwAAQBAJ} {\emph {\bibinfo {title}
  {Molecular Vibrations: The Theory of Infrared and {R}aman Vibrational
  Spectra}}},\ Dover Books on Chemistry Series\ (\bibinfo  {publisher} {Dover
  Publications},\ \bibinfo {address} {New York},\ \bibinfo {year}
  {1980})\BibitemShut {NoStop}%
\bibitem [{Note5()}]{Note5}%
  \BibitemOpen
  \bibinfo {note} {Derivatives with respect to the internal quasicentroid
  coordinates (e.g., $R_{1,2}$ and $\Theta $) are easily expressed in terms of
  the Cartesian bead coordinates, as are the derivatives with respect to the
  molecular centres of mass. However application of the chain rule to the
  remaining external quasicentroid coordinates would require one to derive a $3
  n (N-1)$-dimensional set of curvilinear coordinates orthogonal to $\protect
  \ensuremath {\protect \overline {\protect \ensuremath {\protect \mathbf
  {Q}}}}{}^{(\alpha )}$.}\BibitemShut {Stop}%
\bibitem [{Note6()}]{Note6}%
  \BibitemOpen
  \bibinfo {note} {Derivatives with respect to the Cartesian centroids are
  easily expressed in terms of bead coordinates.}\BibitemShut {Stop}%
\bibitem [{Note7()}]{Note7}%
  \BibitemOpen
  \bibinfo {note} {The ensembles of Matsubara trajectories that survive the
  quasicentroid-constrained Boltzmann averaging are thus expected to be as
  compact as the quasicentroid ring-polymer distributions.}\BibitemShut {Stop}%
\bibitem [{Note8()}]{Note8}%
  \BibitemOpen
  \bibinfo {note} {Here we take $N$ to be even.}\BibitemShut {Stop}%
\bibitem [{\citenamefont {Kapil}\ \emph {et~al.}(2019)\citenamefont {Kapil},
  \citenamefont {Rossi}, \citenamefont {Marsalek}, \citenamefont {Petraglia},
  \citenamefont {Litman}, \citenamefont {Spura}, \citenamefont {Cheng},
  \citenamefont {Cuzzocrea}, \citenamefont {Mei{\ss}ner}, \citenamefont
  {Wilkins}, \citenamefont {Helfrecht}, \citenamefont {Juda}, \citenamefont
  {Bienvenue}, \citenamefont {Fang}, \citenamefont {Kessler}, \citenamefont
  {Poltavsky}, \citenamefont {Vandenbrande}, \citenamefont {Wieme},
  \citenamefont {Corminboeuf}, \citenamefont {K{\"{u}}hne}, \citenamefont
  {Manolopoulos}, \citenamefont {Markland}, \citenamefont {Richardson},
  \citenamefont {Tkatchenko}, \citenamefont {Tribello}, \citenamefont {{Van
  Speybroeck}},\ and\ \citenamefont {Ceriotti}}]{Kapil2019}%
  \BibitemOpen
  \bibfield  {author} {\bibinfo {author} {\bibfnamefont {V.}~\bibnamefont
  {Kapil}}, \bibinfo {author} {\bibfnamefont {M.}~\bibnamefont {Rossi}},
  \bibinfo {author} {\bibfnamefont {O.}~\bibnamefont {Marsalek}}, \bibinfo
  {author} {\bibfnamefont {R.}~\bibnamefont {Petraglia}}, \bibinfo {author}
  {\bibfnamefont {Y.}~\bibnamefont {Litman}}, \bibinfo {author} {\bibfnamefont
  {T.}~\bibnamefont {Spura}}, \bibinfo {author} {\bibfnamefont
  {B.}~\bibnamefont {Cheng}}, \bibinfo {author} {\bibfnamefont
  {A.}~\bibnamefont {Cuzzocrea}}, \bibinfo {author} {\bibfnamefont {R.~H.}\
  \bibnamefont {Mei{\ss}ner}}, \bibinfo {author} {\bibfnamefont {D.~M.}\
  \bibnamefont {Wilkins}}, \bibinfo {author} {\bibfnamefont {B.~A.}\
  \bibnamefont {Helfrecht}}, \bibinfo {author} {\bibfnamefont {P.}~\bibnamefont
  {Juda}}, \bibinfo {author} {\bibfnamefont {S.~P.}\ \bibnamefont {Bienvenue}},
  \bibinfo {author} {\bibfnamefont {W.}~\bibnamefont {Fang}}, \bibinfo {author}
  {\bibfnamefont {J.}~\bibnamefont {Kessler}}, \bibinfo {author} {\bibfnamefont
  {I.}~\bibnamefont {Poltavsky}}, \bibinfo {author} {\bibfnamefont
  {S.}~\bibnamefont {Vandenbrande}}, \bibinfo {author} {\bibfnamefont
  {J.}~\bibnamefont {Wieme}}, \bibinfo {author} {\bibfnamefont
  {C.}~\bibnamefont {Corminboeuf}}, \bibinfo {author} {\bibfnamefont {T.~D.}\
  \bibnamefont {K{\"{u}}hne}}, \bibinfo {author} {\bibfnamefont {D.~E.}\
  \bibnamefont {Manolopoulos}}, \bibinfo {author} {\bibfnamefont {T.~E.}\
  \bibnamefont {Markland}}, \bibinfo {author} {\bibfnamefont {J.~O.}\
  \bibnamefont {Richardson}}, \bibinfo {author} {\bibfnamefont
  {A.}~\bibnamefont {Tkatchenko}}, \bibinfo {author} {\bibfnamefont {G.~A.}\
  \bibnamefont {Tribello}}, \bibinfo {author} {\bibfnamefont {V.}~\bibnamefont
  {{Van Speybroeck}}},\ and\ \bibinfo {author} {\bibfnamefont {M.}~\bibnamefont
  {Ceriotti}},\ }\bibfield  {title} {\enquote {\bibinfo {title} {{i-PI 2.0: A
  universal force engine for advanced molecular simulations}},}\ }\href
  {https://doi.org/10.1016/j.cpc.2018.09.020} {\bibfield  {journal} {\bibinfo
  {journal} {Comput. Phys. Commun.}\ }\textbf {\bibinfo {volume} {236}},\
  \bibinfo {pages} {214} (\bibinfo {year} {2019})}\BibitemShut {NoStop}%
\bibitem [{Note9()}]{Note9}%
  \BibitemOpen
  \bibinfo {note} {This option is invoked in i-PI by setting the \protect
  \texttt {style} attribute of normal-mode frequencies to \protect \texttt
  {wmax-cmd}}\BibitemShut {NoStop}%
\bibitem [{Note10()}]{Note10}%
  \BibitemOpen
  \bibinfo {note} {The value of $\Omega _\protect \text {ref}$ is chosen to
  give a reasonably flat distribution of $\Omega _n$ over the frequency range
  of interest: there is no need to tune it to any characteristic frequency in
  the infrared spectrum.}\BibitemShut {Stop}%
\bibitem [{\citenamefont {Leimkuhler}\ and\ \citenamefont
  {Matthews}(2012)}]{Leimkuhler2012}%
  \BibitemOpen
  \bibfield  {author} {\bibinfo {author} {\bibfnamefont {B.}~\bibnamefont
  {Leimkuhler}}\ and\ \bibinfo {author} {\bibfnamefont {C.}~\bibnamefont
  {Matthews}},\ }\bibfield  {title} {\enquote {\bibinfo {title} {Rational
  construction of stochastic numerical methods for molecular sampling},}\
  }\href {https://doi.org/10.1093/amrx/abs010} {\bibfield  {journal} {\bibinfo
  {journal} {Appl. Math. Res. eXpress}\ }\textbf {\bibinfo {volume} {2013}},\
  \bibinfo {pages} {34} (\bibinfo {year} {2012})}\BibitemShut {NoStop}%
\bibitem [{\citenamefont {Leimkuhler}\ and\ \citenamefont
  {Matthews}(2013)}]{Leimkuhler2013}%
  \BibitemOpen
  \bibfield  {author} {\bibinfo {author} {\bibfnamefont {B.}~\bibnamefont
  {Leimkuhler}}\ and\ \bibinfo {author} {\bibfnamefont {C.}~\bibnamefont
  {Matthews}},\ }\bibfield  {title} {\enquote {\bibinfo {title} {Robust and
  efficient configurational molecular sampling via {L}angevin dynamics},}\
  }\href {https://doi.org/10.1063/1.4802990} {\bibfield  {journal} {\bibinfo
  {journal} {J. Chem. Phys.}\ }\textbf {\bibinfo {volume} {138}},\ \bibinfo
  {pages} {174102} (\bibinfo {year} {2013})}\BibitemShut {NoStop}%
\bibitem [{\citenamefont {Leimkuhler}\ and\ \citenamefont
  {Matthews}(2016)}]{Leimkuhler2016}%
  \BibitemOpen
  \bibfield  {author} {\bibinfo {author} {\bibfnamefont {B.}~\bibnamefont
  {Leimkuhler}}\ and\ \bibinfo {author} {\bibfnamefont {C.}~\bibnamefont
  {Matthews}},\ }\bibfield  {title} {\enquote {\bibinfo {title} {Efficient
  molecular dynamics using geodesic integration and solvent--solute
  splitting},}\ }\href {https://doi.org/10.1098/rspa.2016.0138} {\bibfield
  {journal} {\bibinfo  {journal} {Proc. R. Soc. A Math. Phys. Eng. Sci.}\
  }\textbf {\bibinfo {volume} {472}},\ \bibinfo {pages} {20160138} (\bibinfo
  {year} {2016})}\BibitemShut {NoStop}%
\bibitem [{\citenamefont {Bussi}\ and\ \citenamefont
  {Parrinello}(2008)}]{Bussi2008}%
  \BibitemOpen
  \bibfield  {author} {\bibinfo {author} {\bibfnamefont {G.}~\bibnamefont
  {Bussi}}\ and\ \bibinfo {author} {\bibfnamefont {M.}~\bibnamefont
  {Parrinello}},\ }\bibfield  {title} {\enquote {\bibinfo {title} {Stochastic
  thermostats: comparison of local and global schemes},}\ }\href
  {https://doi.org/10.1016/j.cpc.2008.01.006} {\bibfield  {journal} {\bibinfo
  {journal} {Comput. Phys. Commun.}\ }\textbf {\bibinfo {volume} {179}},\
  \bibinfo {pages} {26} (\bibinfo {year} {2008})}\BibitemShut {NoStop}%
\bibitem [{\citenamefont {Yu}\ and\ \citenamefont {Bowman}(2019)}]{Yu2019}%
  \BibitemOpen
  \bibfield  {author} {\bibinfo {author} {\bibfnamefont {Q.}~\bibnamefont
  {Yu}}\ and\ \bibinfo {author} {\bibfnamefont {J.~M.}\ \bibnamefont
  {Bowman}},\ }\bibfield  {title} {\enquote {\bibinfo {title} {Classical,
  thermostated ring polymer, and quantum {VSCF/VCI} calculations of {IR}
  spectra of {H$_7$O$_3${}$^+$} and {H$_9$O$_4${}$^+$}({Eigen}) and comparison
  with experiment},}\ }\href
  {https://pubs.acs.org/doi/10.1021/acs.jpca.8b11603} {\bibfield  {journal}
  {\bibinfo  {journal} {J. Phys. Chem. A}\ }\textbf {\bibinfo {volume} {123}},\
  \bibinfo {pages} {1399} (\bibinfo {year} {2019})}\BibitemShut {NoStop}%
\bibitem [{\citenamefont {Musil}\ \emph {et~al.}(2022)\citenamefont {Musil},
  \citenamefont {Zaporozhets}, \citenamefont {No\'{e}}, \citenamefont
  {Clementi},\ and\ \citenamefont {Kapil}}]{Venkat2022}%
  \BibitemOpen
  \bibfield  {author} {\bibinfo {author} {\bibfnamefont {F.}~\bibnamefont
  {Musil}}, \bibinfo {author} {\bibfnamefont {I.}~\bibnamefont {Zaporozhets}},
  \bibinfo {author} {\bibfnamefont {F.}~\bibnamefont {No\'{e}}}, \bibinfo
  {author} {\bibfnamefont {C.}~\bibnamefont {Clementi}},\ and\ \bibinfo
  {author} {\bibfnamefont {V.}~\bibnamefont {Kapil}},\ }\href@noop {} {\enquote
  {\bibinfo {title} {Quantum dynamics using path integral coarse-graining},}\ }
  (\bibinfo {year} {2022}),\ \Eprint {https://arxiv.org/abs/2208.06205}
  {arXiv:2208.06205} \BibitemShut {NoStop}%
\bibitem [{\citenamefont {Babin}, \citenamefont {Medders},\ and\ \citenamefont
  {Paesani}(2014)}]{Babin2014}%
  \BibitemOpen
  \bibfield  {author} {\bibinfo {author} {\bibfnamefont {V.}~\bibnamefont
  {Babin}}, \bibinfo {author} {\bibfnamefont {G.~R.}\ \bibnamefont {Medders}},\
  and\ \bibinfo {author} {\bibfnamefont {F.}~\bibnamefont {Paesani}},\
  }\bibfield  {title} {\enquote {\bibinfo {title} {Development of a ``first
  principles'' water potential with flexible monomers. {II}: {T}rimer potential
  energy surface, third virial coefficient, and small clusters},}\ }\href
  {http://pubs.acs.org/doi/10.1021/ct500079y} {\bibfield  {journal} {\bibinfo
  {journal} {J. Chem. Theory and Comput.}\ }\textbf {\bibinfo {volume} {10}},\
  \bibinfo {pages} {1599} (\bibinfo {year} {2014})}\BibitemShut {NoStop}%
\bibitem [{\citenamefont {Marsalek}\ and\ \citenamefont
  {Markland}(2017)}]{Marsalek2017}%
  \BibitemOpen
  \bibfield  {author} {\bibinfo {author} {\bibfnamefont {O.}~\bibnamefont
  {Marsalek}}\ and\ \bibinfo {author} {\bibfnamefont {T.~E.}\ \bibnamefont
  {Markland}},\ }\bibfield  {title} {\enquote {\bibinfo {title} {Quantum
  dynamics and spectroscopy of ab initio liquid water: The interplay of nuclear
  and electronic quantum effects},}\ }\href
  {https://doi.org/10.1021/acs.jpclett.7b00391} {\bibfield  {journal} {\bibinfo
   {journal} {J. Phys. Chem. Lett.}\ }\textbf {\bibinfo {volume} {8}},\
  \bibinfo {pages} {1545} (\bibinfo {year} {2017})}\BibitemShut {NoStop}%
\end{thebibliography}%

\end{document}